\documentclass[twocolumn,english,aps,prl,superscriptaddress]{revtex4-1}
\usepackage{amsmath}
\usepackage{amssymb}
\usepackage{graphicx}
\usepackage{esint}
\usepackage{xcolor}
\usepackage{mathtools}
\newcommand{\kk}{\vec{k}} 
\usepackage[colorlinks=true]{hyperref}
\usepackage{ulem}
\hypersetup{
	bookmarks=true,         
	unicode=false,          
	pdftoolbar=true,        
	pdfmenubar=true,        
	pdffitwindow=false,     
	pdfstartview={FitH},    
	pdftitle={S},    
	pdfauthor={S. Sachdev},     
	pdfsubject={},   
	pdfcreator={},   
	pdfproducer={}, 
	pdfkeywords={} {} {}, 
	pdfnewwindow=true,      
	colorlinks=true,       
	linkcolor=magenta, 
	citecolor=blue,        
	filecolor=magenta,      
	urlcolor=blue           
}

\makeatletter

\usepackage{babel}

\usepackage{hyperref}

\makeatother

\usepackage{babel}
\begin{document}
\title{Thermal Hall conductivity near field-suppressed magnetic order in a Kitaev-Heisenberg model}

\author{Aman Kumar}

\affiliation{Department of Theoretical Physics, Tata Institute of Fundamental
Research, Homi Bhabha Road, Navy Nagar, Mumbai 400005, India}
\affiliation{Department of Physics, Harvard University, Cambridge MA-02138, USA}

\author{Vikram Tripathi}

\affiliation{Department of Theoretical Physics, Tata Institute of Fundamental
Research, Homi Bhabha Road, Navy Nagar, Mumbai 400005, India}

\date{\today}
\begin{abstract}
We investigate thermal Hall conductivity $\kappa_{xy}$ of a $J$-$K$ Kitaev-Heisenberg model with a Zeeman field in the $(111)$ direction in the light of the recent debate surrounding the possible re-emergence of Ising topological order (ITO) and half-quantized $\kappa_{xy}/T$ upon field-suppression of long-range magnetic order in Kitaev materials. We use the purification-based finite temperature Tensor Network approach making no prior assumptions about the nature of the excitations: Majorana, visons or spin waves. For purely Kitaev interactions and fields $h/K \gtrsim 0.02$ sufficient to degrade ITO, the peak $\kappa_{xy}/T$ monotonously decreases from half-quantization associated with lower fields - a behavior reminiscent of vison fluctuation corrections. For higher fields $h/K\gtrsim 0.1,$ we find the results qualitatively consistent with a spin-wave treatment. In our $J$-$K$ model (with ferro-$K$ and antiferro-$J$), in the vicinity of field-suppressed magnetic order, we found $\kappa_{xy}/T$ to be significant, with peak magnitudes exceeding half-quantization followed by a monotonous decrease with increasing $h.$ We thus conclude that half-quantized thermal Hall effect in the vicinity of field suppressed magnetic order in our model, is a fine-tuning effect and is not associated with a Majorana Hall state with ITO. 
\end{abstract}
\maketitle

Thermal Hall conductivity has become a very important transport diagnostic for emergent phenomena in insulating magnetic systems such as the underdoped cuprates \cite{grissonnanche2020chiral}, frustrated magnets such as pyrochlores \cite{hirschberger2015thermal_pyrochlore}, kagome systems \cite{hirschberger2015thermal_kagome}, and more recently a Kitaev material $\alpha$-RuCl$_3$ \cite{kasahara2018unusual,yokoi2021half,bruin2021robustness,czajka2021oscillations}. Specifically in the Kitaev material, one set of studies \cite{yokoi2021half,kasahara2018unusual,bruin2021robustness} has reported the half-quantized thermal Hall effect at finite magnetic fields comparable to the magnetic ordering scale, suggesting a re-emergent Majorana Hall insulator state originally predicted \cite{kitaev2006anyons} for the honeycomb Kitaev model at very small magnetic fields and low temperatures, which is a deconfined phase with robust Ising topological order (ITO). This view has been challenged \cite{czajka2021oscillations,czajka2023planar} in recent experiments where half-quantization has been ruled out. Nevertheless $\kappa_{xy}/T$ is found to be large, sometimes even exceeding the half-quantized value at temperatures and fields comparable to the magnetic ordering scale, implying half-quantized peaks/plateaus of $\kappa_{xy}/T,$ if observed near field suppressed magnetic order, are a consequence of fine tuning and do not represent re-emergence of a deconfined ITO phase. The authors of Refs. \cite{czajka2021oscillations,czajka2023planar} offered an alternate explanation based on a spin wave picture. Theoretical studies have relied on various interacting quasiparticle pictures, with spin wave approximations \cite{Kim_sign_structure,zhang2021topological,koyama2021field,mcclarty2018topological,Joshi_prb_magnon} most commonly used. Since spin wave approximations in frustrated spin-$1/2$ magnets are best justified at high fields or in the presence of long-range magnetic order, the possibility of half-quantization at low temperatures near field-suppressed magnetic order (not seen in spin wave treatments) cannot be completely ruled out. Other theoretical approaches begin from the fractionalized quasiparticles, and suppression of half-quantized $\kappa_{xy}/T$ by external magnetic field or finite temperatures is understood from the perspective of vison (gauge field) fluctuations that couple to the Kitaev spinons \cite{guo2020gauge}, or from thermal excitation of bulk Majorana fermions \cite{nasu2017thermal}. Although the fractionalization analyses are based around expansion from the ITO phase, even here it is not clear whether ITO can re-emerge through field-suppression of long-range magnetic order. 
These challenges motivated us to take a different route of numerically estimating $\kappa_{xy}$ without making \emph{any} quasiparticle approximation - whether fractionalized or spin waves - which forms the basis of existing theoretical approaches. 
We obtain the thermal Hall current and conductivity numerically using purification-based Tensor Network techniques\cite{purification_pollmann}. To our knowledge, this is the first use of the Tensor Network technique for thermal Hall calculation of an interacting many-body system.
 
 We consider for concreteness a ferromagnetic (FM) Kitaev model with a competing antiferromagnetic (AFM) Heisenberg interaction ($J$-$K$ model), subjected to a Zeeman field along the $\mathbf{c} = (1,1,1)$ direction,
 \begin{align}
  H & = - K\!\!\!\!\! \sum_{\langle ij\rangle;\gamma-\text{links}}\!\!\!\!\!S_{i}^{\gamma}S_{j}^{\gamma} + J \sum_{\langle ij\rangle}\mathbf{S}_{i}\cdot\mathbf{S}_{j} + h \sum_{i,\gamma}S_{i}^{\gamma}.
  \label{eq:model}
 \end{align}
 Here  $\gamma = x, y, z$ denote the bond dependent spin axes on the honeycomb lattice. Thermal Hall conductivity of this model has recently been studied using spin wave approximations \cite{koyama2021field}. In the Kitaev limit ($J/K=0$), there is a fairly large window of magnetic fields (e.g. $h_{\parallel \mathbf{c}} \gtrsim 0.02 K$) \cite{Kitaev_field_gap,Gohlke_dynamics_kitaev} where ITO is significantly degraded for a similar signature in the topological entanglement entropy $\gamma(h)$). Interacting spin wave analysis \cite{mcclarty2018topological} shows that magnons are good quasiparticles at high fields. Fields in the range $0.02 \lesssim h/K \lesssim 1$ constitute an interpolating region, and we focus on this window. Likewise for the spin density wave (SDW) ordered states of the $J$-$K$ model, we are interested in the range of magnetic fields starting from field-suppressed SDW to deeper in the spin-polarized phase, where the nature of excitations may be changing.  
 
Our main findings are as follows. In the Kitaev limit ($J/K=0$), at the lower end of our field range $h/K= 0.025,$ which lies on the boundary of the ITO phase, $\kappa_{xy}/T$ peaks close to the half-quantized value $\pi/12$ associated with the ITO phase (see Fig. \ref{fig:thermalhall}). With increasing $h$ as well as temperature $T,$ $\kappa_{xy}/T$ declines sharply and monotonously, reminiscent of vison fluctuation effects discussed recently \cite{guo2020gauge}. The fall of the peak height mirrors a similar sharp fall in the toplogical entanglement entropy $\gamma$ of the ground state beyond $h/K \approx 0.02,$ from the ITO value of $\ln 2$ for $0\leq h/K \lesssim 0.02.$ The small $\gamma$ for $h/K \gtrsim 0.1$ suggests the spinon approach might not be a good starting point at higher fields, and we compare with the spin-wave approximation here. We used the formalism developed in Refs. \cite{guo_thermal,kapustin_thermal,qin_phonon} and found qualitative agreement with our numerical results. 

In the magnetically ordered phase $\gamma$ takes small values signifying absence of ITO throughout the field range. Nevertheless $\kappa_{xy}/T$ peaks in the region of field-suppressed magnetic order, with an opposite sign relative to the Kitaev limit, and peak values that even exceed half-quantization, reminiscent of experimental reports \cite{czajka2023planar}. Further increasing the magnetic field decreases the extrema, which monotonously approach zero at high fields like in the Kitaev limit. We posit that half-quantized thermal Hall effect appears in the ITO phase and possibly its immediate vicinity, i.e., at low fields and no magnetic order. Therefore in the field-suppressed SDW regime, we do not accord physical significance to any isomagnetic of $\kappa_{xy}/T$ showing extremum near half-quantization since the isomagnetics can peak at much larger values for nearby fields. 
 
 \emph{Strategy}: Thermal currents can be readily expressed in terms of local microscopic degrees of freedom (spin operators in our case) and do not require any knowledge of quasiparticles, which was the approach followed in Kitaev's original work \cite{kitaev2006anyons}. Kitaev's prescription for calculation of edge thermal currents is however difficult to implement numerically in a general nonintegrable situation because of the large temperature  differences used to mimic the bulk and vacuum regions. An alternate way of using the Kubo linear response approach is also numerically challenging because it involves small differences of two large contributions represented by the energy-current correlators and the bound energy magnetization contributions \cite{Cooperthermoelectric}. Our strategy is similar to Kitaev's except that the system is immersed in a uniform temperature bath. Although there is no net transport Hall current, it is because of cancellation of equal and opposite contributions of the two edges. Then, for gapped systems that are sufficiently long, the thermal Hall current associated with one edge is simply obtained by summing over the Hall currents from the relevant edge to the center. This method assumes the thermal Hall currents are essentially due to edge modes and the Hall current deep inside the bulk is insignificant; hence the decay of thermal Hall currents away from the bulk needs to be checked every time. This also ensures that the left and right edge currents do not overlap.
 
 \begin{figure}
 	\includegraphics[width=1\linewidth]{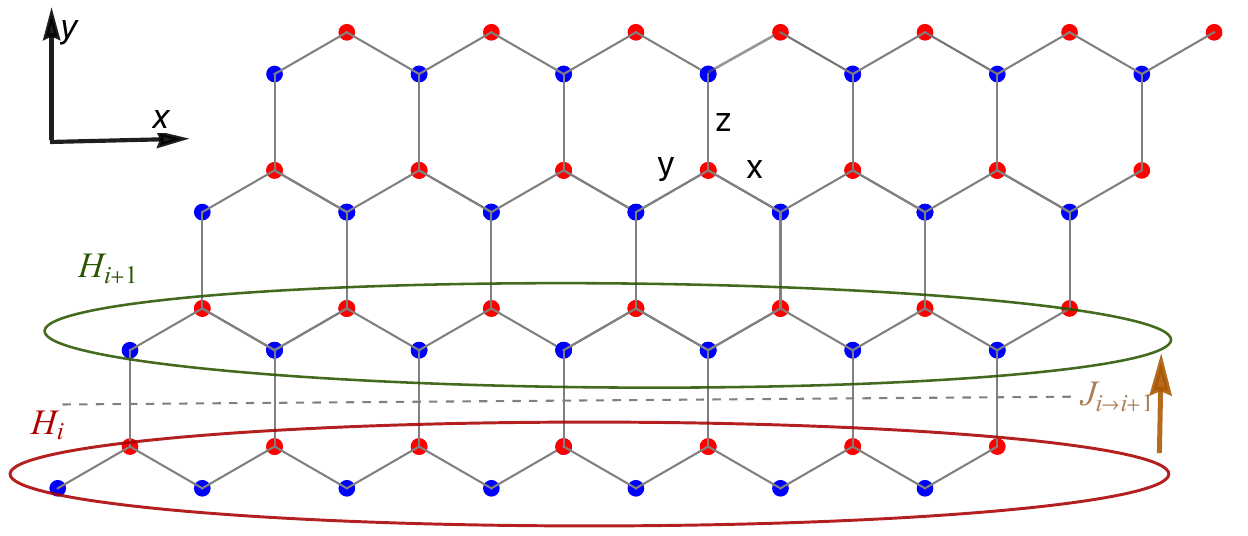}
 	\caption{\label{fig:schematic}Cylindrical geometry for the honeycomb Kitaev model used in our calculations, with periodic boundary conditions along the $y$-direction and open zigzag edges with $L_y$ unit cells. Calculations have been performed with $N_{y}=2L_{y}=6$ of sites along the $y$ direction and convergence checked for $N_{y}=8.$ There are up to $N_{x}=L_{x}=20$ sites along the $x$-direction, the cylinder's axis. The dashed dual lattice lines separate adjacent layers (ellipsoids) associated with Hamiltonian $H_{i}$, $H_{i+1}.$}
 \end{figure}
 
 For calculations, we put our model on a cylinder along the $x$-direction with periodic boundary conditions along the $y$-direction, with a pair of open zigzag edges, see Fig.~\ref{fig:schematic}. The Hamiltonian is re-expressed as a sum of layer contributions, $H=\sum_{i}H_{i},$
connecting the two open edges. Links that cut the dual curve (dashed line in Fig.~\ref{fig:schematic}) separating adjacent layers, make equal contributions to the adjacent $H_{i}.$ The energy current across the dual curve is obtained from the continuity equation,
\begin{align}
\frac{dH_{i}}{dt}\,=\, -i[H_{i},H]\,=\,\sum_{j}\hat{J}_{ij},
 \quad \hat{J}_{ij}=-i[H_{i},H_{j}].
\end{align}
Here $\hat{J}_{ij}$ is the energy current from layer $j$ to $i,$ and we consider it as a sum of contributions $\hat{J}_{ij}(x)$ starting from one edge ($x=0$) to the other ($x=L_x$). In situations where the thermal Hall currents are essentially due to the edge modes, the Hall current associated with one of the edges is given by $J_{ij}^{\text{H}}=\sum_{x\leq L_{x}/2} \langle \hat{J}_{ij}(x)\rangle.$ The averaging includes both quantum and thermal. In the steady state, the labels $ij$ can also be dropped.
Thermal Hall conductivity $\kappa_{xy}$ is obtained by taking the temperature derivative of $J^{\text{H}}.$

We use the standard purification-based finite temperature Tensor Network method \cite{purification_pollmann} for performing the quantum and thermal averaging, where imaginary time evolution of the state is carried by applying the evolution operator as a matrix product operator (MPO) with W-II approximation, described in Ref. \cite{time_evolution_zaletel} -- see appendix for more details.
Benchmarking of the finite temperature purification method was done (see appendix) against Exact Diagonalization (ED) calculation of one of the Kitaev plaquette fluxes at finite temperature and a small field of $h/K=0.025,$ and we found very good agreement between the two even at temperatures as low as $T/K=0.01.$ However the thermal Hall calculations at even smaller fields taking us well inside the ITO phase are substantially more expensive to implement numerically because of higher bond dimension and smaller Trotter size requirements. We complement the finite temperature study of $\kappa_{xy}/T$ with finite DMRG calculations of ground state topological entanglement entropy $\gamma$ both inside and outside the ITO phases. Note that the ITO phase is rather well-studied in the literature, and moreover the experimental interest is in magnetic fields that lie well outside the ITO  - this further motivates us to focus in the field range upwards of those needed to suppress ITO.
%

\emph{Results}: We consider first $\kappa_{xy}/T$ in the Kitaev limit $J/K=0,$  see Fig. \ref{fig:thermalhall}. Introduction of a magnetic field hybridizes the dispersing majoranas with pairs of visons \cite{zhang2022theory} whose excitation energy at zero field is $\Delta_{\text{pair}}\approx 0.065 K.$ The hybridization results in gapping of the bulk majoranas and appearance of chiral Majorana edge modes \cite{kitaev2006anyons}, vison fluctuations/hopping, and a suppression of topological order \cite{purification_pollmann,Kitaev_field_gap} beyond $h\sim 0.02 K$ with a concomitant gapping of the edge states \cite{Kitaev_field_gap}. Fractionalization may however survive to higher fields, as evidenced from quasiparticle stability analysis \cite{Aman2020}, and also experimentally from specific heat measurements \cite{tanaka2022thermodynamic}. At high fields $h/K \gtrsim 0.1,$ fractionalization is lost (evidenced e.g. in the steep fall in $\gamma,$ see inset of Fig. \ref{fig:thermalhall}) and local spin flips (magnons) constitute the elementary excitations.    

Figure \ref{fig:thermalhall} shows the temperature dependence of $\kappa_{xy}/T$ in the Kitaev limit for different values of $h.$  The temperatures corresponding  to the peak values of $\kappa_{xy}/T$ all lie below the bulk gap $\Delta_{\text{bulk}}$\footnote{The bulk gap is calculated using DMRG on a torus for a 32-site system.}, see table in appendix. Thus the low temperature peaks are associated with gapped edge modes. This is also directly confirmed by our observation of rapidly decaying thermal Hall currents away from the edge and into the bulk (see appendix). For all fields  $h/K>0.03,$ we found the peak temperature is not very sensitive to the finite circumference of the edge - we confirmed this by increasing the edge size from $N_{y}=6$ to $N_{y}=8$ spins. An edge gap is known to appear upon transitioning out of the ITO phase \cite{Kitaev_field_gap}. For the lowest field shown in Fig. \ref{fig:thermalhall}, $h/K=0.025,$ the peak value of $\kappa_{xy}/T$  is close to the half-quantized value associated with the ITO phase. Further increase of $h$ monotonously decreases the peak values, while simultaneously pushing the peak to higher temperatures. At high temperatures, the $\kappa_{xy}/T$ all tend to vanish and we see no signs of saturation predicted from an earlier spin wave \cite{mcclarty2018topological} calculation, and we revisit the spin-wave treatment below. We checked that $\kappa_{xy}/T$ has a similar behavior (although with an opposite sign) for fields along the crystallographic $\mathbf{a}=(1,1,\bar{2})$ direction that is often the case experimentally. The sign of $\kappa_{xy}/T$ is dictated by that of the product $h_x h_y h_z$ as shown in Kitaev's original work. We also verified that the effect vanishes along the $\mathbf{b}=(1,\bar{1},0)$ direction - a consequence of the so-called $R^{*}$-symmetry \cite{koyama2021field,guo2020gauge}.

\begin{figure}
\includegraphics[width=1\columnwidth]{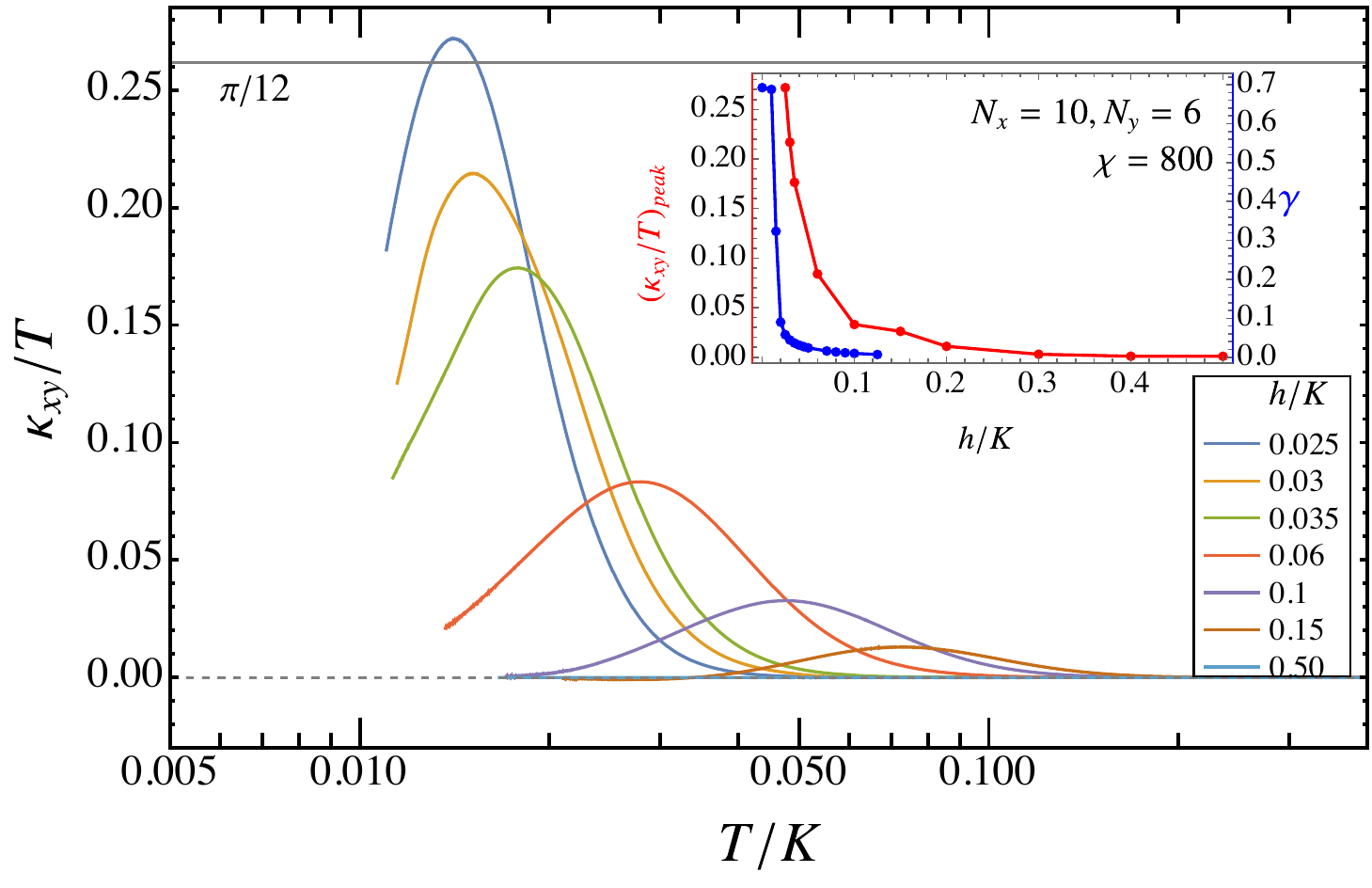}
\caption{\label{fig:thermalhall}Isomagnetic plots of $\kappa_{xy}/T$ for the ferromagnetic Kitaev model in the field range $0.025 \leq h/K\leq 0.5$ interpolating the Ising topological order (ITO) and spin wave limits. The chosen system size $N_x=10$ was enough for Hall current to decay in bulk. At low fields, signatures of proximate ITO are visible in the form of $\kappa_{xy}/T$ peaks that approach half-quantization at the lower end. Thermal Hall effect for this range of fields appears more consistent with a picture of interacting fractionalized quasiparticles \cite{guo2020gauge} than spin waves \cite{mcclarty2018topological}. Solid gray line describes the half-quantized thermal Hall effect. Inset shows the plot of peaked $\kappa_{xy}/T$ versus magnetic field on the left $y$ axes and topological entanglement entropy $\gamma$ on the right $y$ axes in pure Kitaev limit.}
\end{figure}

\begin{figure}
\includegraphics[width=1\columnwidth]{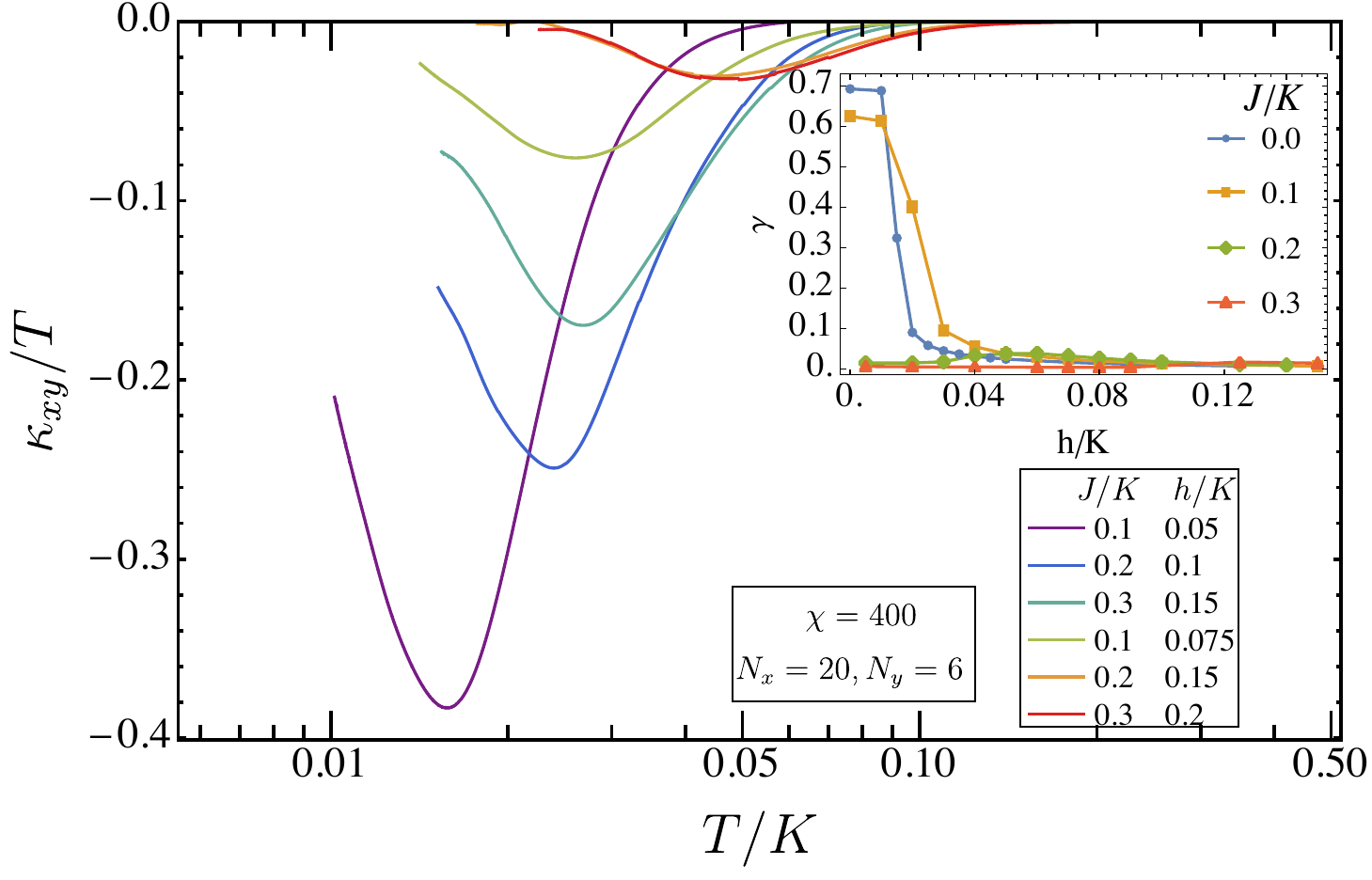} 
\caption{\label{fig:J-K} Isomagnetic plots of $\kappa_{xy}/T$ for the $J$-$K$ model, with Heisenberg interactions chosen both inside the Kitaev spin liquid phase ($J/K=0.1$) as well as in the SDW phase ($J/K=0.2,\,0.3$). For $J/K=0.1,$ the fields chosen exceed the region of ITO order, and for the SDW, the fields are in the vicinity of the partially revived ITO and beyond. Generically, $\kappa_{xy}/T$ is characterized by deep troughs at low fields that get shallower as the field increases. The minimum values in the vicinity of partially revived ITO typically go beyond half-quantization. For the field-suppressed SDW cases, we found no phase characterized by half-quantization and accompanying fully revived ITO. Inset shows the plot  topological entanglement entropy $\gamma$ as a function of magnetic field $h/K$ for different values of $J/K.$}
\end{figure}

We now turn on the competing Heisenberg AFM interaction $J.$ At $h=0,$ the Kitaev spin liquid phase is known to survive up to $J/K \approx 0.1$  \cite{JackeliKitaevHeisenberg,dynamics_gohlke}, beyond which a stripy SDW phase appears. The $(111)$ Zeeman field competes with both topological and magnetic order \cite{trebst_PhysRevB.83.245104}, and at large values yields a trivial polarized phase. 
Recent field-theoretical phenomenology \cite{QCD3_chernPhysRevResearch.2.013072} as well as thermal Hall measurements of $\alpha$-RuCl$_3$ support the view that ITO re-emerges upon field-suppression of SDW order. However other experimental \cite{czajka2021oscillations} and theoretical studies based on spin wave treatments \cite{mcclarty2018topological,koyama2021field} disagree, and even the robustness of the reported half-quantization in this regime has been questioned. Before examining $\kappa_{xy}/T,$ which depends on the excitations in the model, we first checked if the ground state state shows re-emerged ITO by calculating $\gamma(h)$ for different values of $J/K$ (see inset \ref{fig:J-K} and appendix for more details.). For the magnetically ordered phase, $J/K > 0.1,$ we found that $\gamma(h)$ is small at all fields, with a tiny revival near $h\sim J/3$ corresponding to field suppressed magnetic order. In the discussion below we focus our attention in the vicinity in this field regime since $\kappa_{xy}/T$ also peaks here.

Figure \ref{fig:J-K} shows isomagnetic curves of $\kappa_{xy}/T$ for different values of $J/K$ near the field-suppressed stripy AFM regime. Consider the curve corresponding to $(J/K,h/K)=(0.1,0.05).$ Apart from the sign reversal, $\kappa_{xy}/T$ shows a deep minimum that exceeds the half-quantized value, indicating we are well outside the ITO regime at such fields. This is also supported by the strong degradation of $\gamma(h)$ (see inset of Fig. \ref{fig:J-K}) for the above parameters. We noticed that sign reversal of $\kappa_{xy}/T$ happens at $J/K=0.02$, observed from the computation of the edge current in the ground state. Higher values of $J/K =0.2,\, 0.3$ take us well within the stripy SDW phase at $h=0.$ We dial up the field to values where ITO is partially revived, and calculate $\kappa_{xy}/T$ in its vicinity. We find deep minima of comparable strength to the one seen for $(J/K,h/K)=(0.1,0.05).$
 Upon increasing the field, the minimum becomes shallower, very reminiscent of the intermediate field behavior of the Kitaev model where ITO has degraded. The temperature corresponding to the extremum lies well within the bulk gap (see Table in SM \cite{SM}) for these parameter values.
In regimes with strong magnetic order and low fields ($h/J\ll 1$), we found that the Hall currents do not decay in the bulk for the length-scales we could study; consequently restricting us to the field-suppressed SDW regime and beyond. 
This is a limitation of our method that we have mentioned already - in cases where the bulk makes a significant contribution to the thermal Hall currents, our method of calculating the Hall current is invalidated.
However it is encouraging for us that experimentally, the thermal Hall response at low fields in the SDW phase is not observed to be significant \cite{kasahara2018unusual}, and this is also the case in spin wave treatments \cite{koyama2021field} that are relevant for magnetically ordered states.  

\begin{figure}
	\includegraphics[width=1\columnwidth]{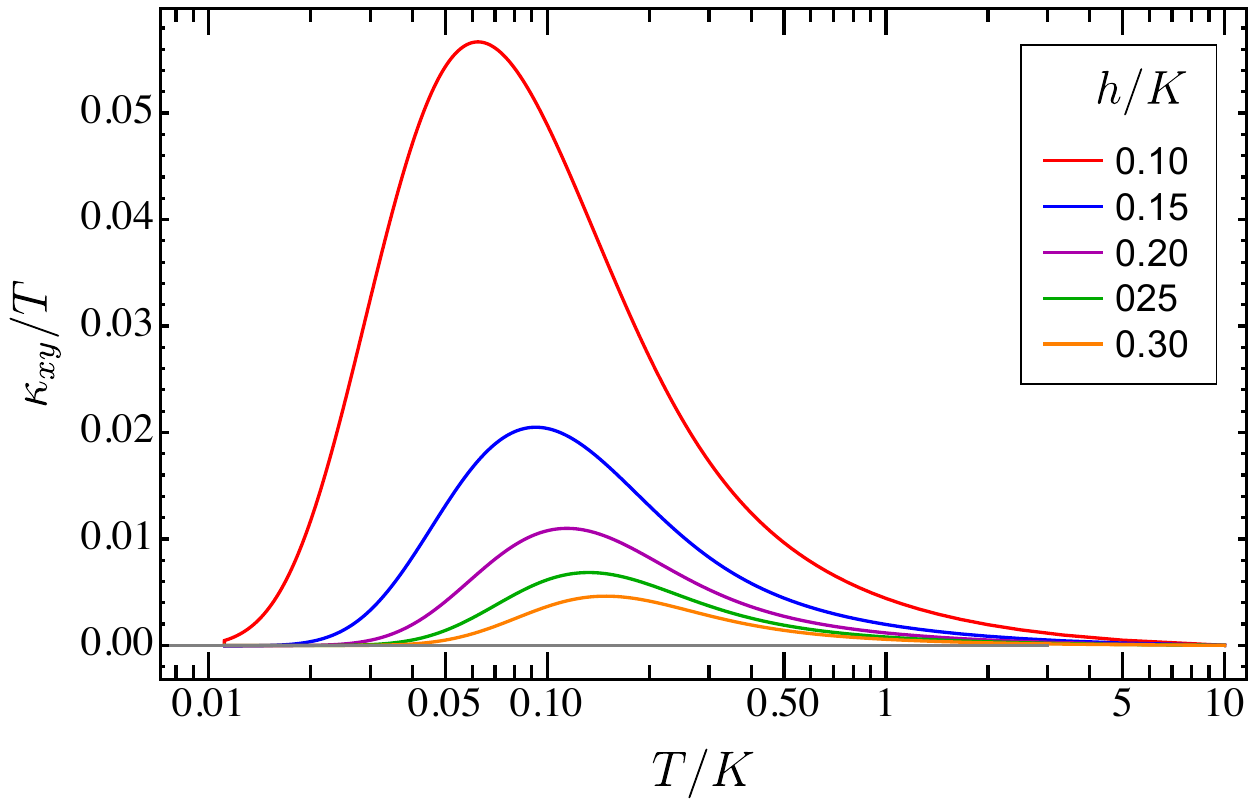}
	\caption{\label{fig:spin_wave} Isomagnetic plots of $\kappa_{xy}/T$ for the Kitaev model in (1,1,1) magnetic field. Curves corresponds to the spin wave theory based results.}
\end{figure}
We conclude with a comparison with spin-wave results for fields $h/K \geq 0.1.$ Our calculations are based on the formalism developed in Refs. \cite{guo_thermal,kapustin_thermal,qin_phonon} which explicitly takes into account the subtraction of bound energy magnetization currents, and differs from the expression obtained in Ref. \cite{murakami_thermal} that have been used in Ref. \cite{mcclarty2018topological}. Figure \ref{fig:spin_wave}  shows the results of our spin-wave calculations (see also appendix). The behavior is qualitatively similar to Fig. 2 but significantly differs from the earlier study\cite{mcclarty2018topological}.

To summarize, we studied the temperature dependence of thermal Hall conductivity of the Kitaev-Heisenberg model for Zeeman fields in the $(111)\parallel \mathbf{c}$ direction using a purification based finite temperature Tensor Network technique that goes down to $K/T \approx 100.$ We make no prior assumption about the nature of the quasiparticles, whether fractionalized ones or spin waves. In the Kitaev limit, we focused on fields $h/K\geq 0.025$ from the boundary of the ITO phase any beyond into the polarized paramagnetic phase. We found that $\kappa_{xy}/T$ features a low-temperature peak whose value approaches half-quantization at the lower end of our field range, and continuously declines with increasing $h,$ approaching zero at larger fields. We associate the low-temperature peak with gapped edge modes expected in the absence of Ising topological order \cite{senthil_2022translation}. After peaking, $\kappa_{xy}/T$ monotonously decreases with increasing temperature reminiscent of vison fluctuation effects \cite{guo2020gauge}. For higher fields $h/K \gtrsim 0.1,$ we made a spin-wave treatment and found $\kappa_{xy}/T$ qualitatively agrees with the tensor network calculations, indicating a smooth crossover from a fermionic spinon picture to a bosonic spin-wave one. The spin-wave calculations are performed using the formalism of Refs. \cite{kapustin_thermal,guo_thermal} correctly account for energy magnetization effects, and the results differ from an earlier one \cite{mcclarty2018topological} that suggests nonvanishing $\kappa_{xy}/T$ at high temperatures.

To understand thermal Hall response of a field-suppressed SDW phase of the Kitaev-Heisenberg model, we considered the cases $J/K=0.1$ at the boundary of the spin liquid phase, and $J/K=0.2,\,0.3$ that lie well within the stripy SDW phase. Characteristically deep minima were seen in the isomagnetic curves of $\kappa_{xy}/T$ versus $T,$ significantly exceeding half-quantization of the ITO phase, similar to the experimental observations in $\alpha$-RuCl$_3;$ however, calculation of topological entanglement entropy showed the absence of ITO here. The sign, which is opposite to the pure Kitaev limit together with magnitudes exceeding half-quantization supports the absence of a Majorana Hall insulator.
 Since the magnitude of $\kappa_{xy}/T$ in the $J$-$K$ model may exceed the half-quantized value, it is possible to fine tune the field to bring it near half-quantization but that does not imply ITO. We believe that similar conclusions should also hold near field-suppressed SDW order even in the presence of other competing interactions like the anisotropic terms $\Gamma, \Gamma'$ - this is ultimately to be decided in future studies.

\begin{acknowledgments}
The authors acknowledge support of the Department of Atomic Energy, Government of India, under Project Identification No. RTI 4002, and the Department of Theoretical Physics, TIFR, for computational resources. They are also grateful to Subir Sachdev, Darshan Joshi and Y.-B. Kim for helpful discussions on the thermal Hall effect. AK thanks Sounak Biswas for discussions on finite temperature tensor network methods.
\end{acknowledgments}
\appendix
\section{Purification method and benchmarking}
\begin{figure}
	\includegraphics[width=1\linewidth]{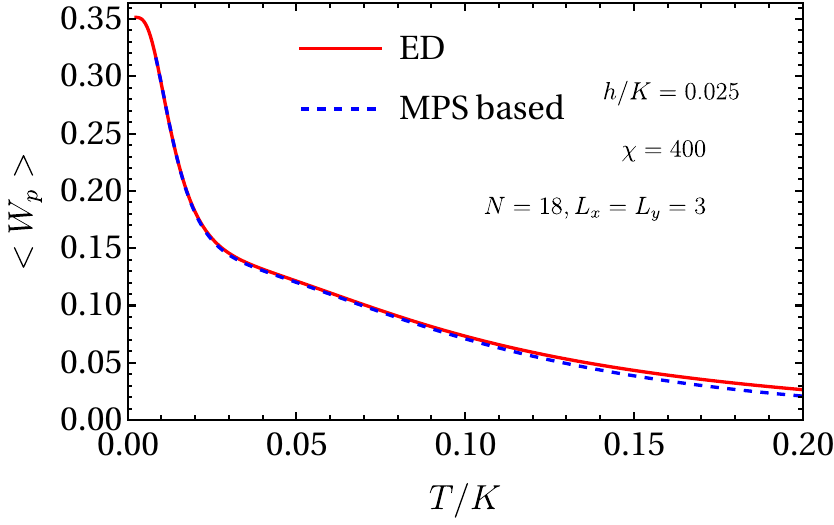}
	\caption{\label{fig:flux}Benchmarking done via calculation of flux value for an $18$-spin system with periodic boundary conditions using ED and MPS based purification methods. The bond dimension $\chi=400$ in the MPS calculation is adequate for good agreement with ED at the low temperature end. At higher temperatures, the small discrepancy is actually on account of ED - the number of excited states used is insufficient for correct description of the high temperature behavior.}
\end{figure}

For calculation of finite temperature quantities, we have used a purification based tensor network method \cite{purification_pollmann}. 
The purification strategy involves expressing the mixed state as a pure state in an enlarged Hilbert space.  We obtain the finite temperature (T=1/$\beta$) thermofield state (TFD) $|\psi_{\beta}\rangle$ by action of imaginary time evolution operator $\exp(-\frac{\beta}{2}H)$ on the infinite temperature state $|\psi_0\rangle$, which is constructed by Bell entangled pair of physical and ancilla spin $1/2$ degree of freedom  at each sites.  Here, $H$ acts on the physical part of Hilbert space and we  use the compact matrix product operator (MPO) representation for the finite temperature evolution operator with  W-II approximation \cite{time_evolution_zaletel}, provided in TenPy library \cite{tenpy}.
Finite temperature expectation of a physical observable $\hat{O}$ is obtained from $\langle \hat{O} \rangle_{\beta}=\langle\psi_{\beta}|\hat{O}|\psi_{\beta}\rangle.$ The purification method works best at higher temperatures. 
Figure \ref{fig:flux} shows the benchmarking of the  method against Exact Diagonalization (ED) calculation of one of the Kitaev plaquette fluxes $\langle W_{p} \rangle$ at finite temperatures and a small field of $h/K=0.025$ for N= 18 site system. We find very good agreement between the two even at temperatures as low as $T/K=0.01.$ At higher temperatures, the discrepancy between purification and ED results is actually on account of the latter being inadequate (insufficient number of excited states taken into consideration).
\section{Additional information on $\kappa_{xy}/T$ calculations}
\begin{figure}
	\includegraphics[width=1\linewidth]{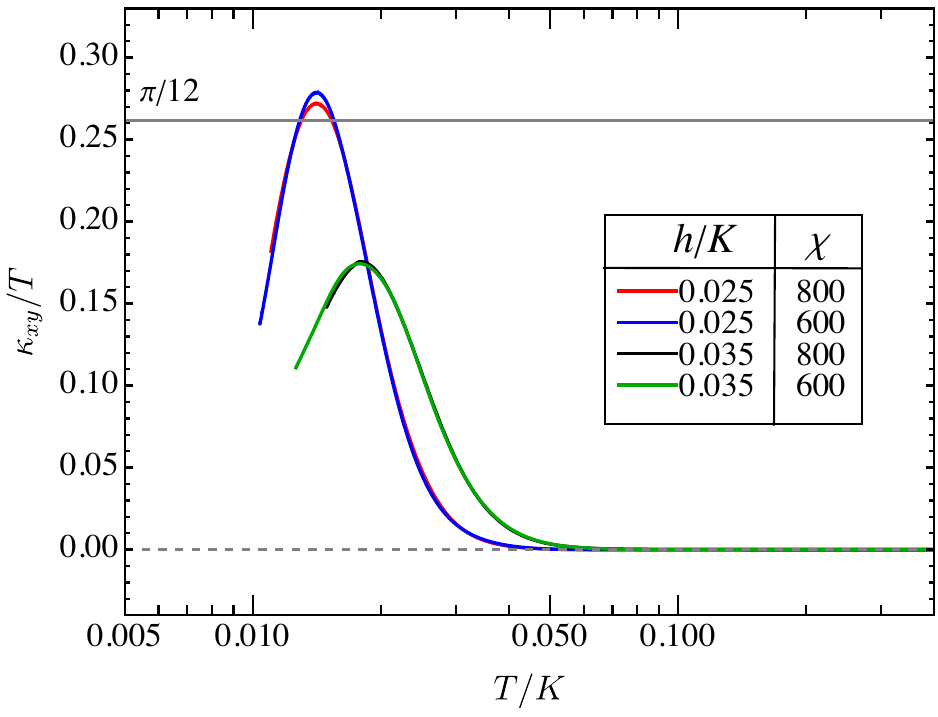}
	\caption{\label{fig:convergence} Plot shows the $\kappa_{xy}/T$ vs. $T/K$ for different bond dimensions $\chi$ in the pure Kitaev limit in two different magnetic fields, with $h/K=0.025$ at the boundary of Ising topological order. $\kappa_{xy}/T$ has converged (wrt increasing $\chi$) for the higher magnetic field, but for field near the boundary of the topolgical phase, it has not converged yet. However the trend of the peak value for $h/K=0.025$ is in the right direction, namely half-quantization.}
\end{figure}

The accuracy of matrix product state (MPS) calculations depends on enhancing the bond dimension $\chi$ and reducing the Trotter step size $\Delta\beta$ used in W-II approximation  \cite{time_evolution_zaletel}. Figure \ref{fig:convergence} shows $\kappa_{xy}/T$  in the Kitaev limit  for two different values of the bond dimension $\chi$. Calculations in and near the boundary of ITO phase typically require larger $\chi$ and smaller $\Delta\beta$. However we have reduced the $\Delta\beta=0.02$ to the lowest of our reach and looked at the convergence with increasing bond dimension. For $h/K=0.025$ which is near the topological phase boundary, the peak of $\kappa_{xy}/T$ overshoots the expected half-quantized value but the overshooting reduces upon increasing the bond dimension. For $h/K=0.035,$ $\kappa_{xy}/T$ has clearly converged with respect to increasing $\chi.$ 

Table \ref{tab:gap} shows the temperature $T^{*}$ at which $\kappa_{xy}/T$ peaks, which we find lies well with in the bulk gap $\Delta_{\text{bulk}}$. This indicates dominance of edge modes to the thermal Hall response in our model. 
\\

\begin{table}
	\begin{ruledtabular}
		\begin{tabular}{llll}
			$(J/K,h/K)$	&$T^{*}/K$  &$\Delta_{\text{bulk}}/K$  \\ \hline
			(0,0.03)&0.0149&0.0473\\
			(0, 0.035) &0.0182&0.0494  \\ 
			(0,  0.06) &0.0278&0.0720  \\
			(0, 0.10) &	0.0482&0.1160 \\ 
			(0, 0.15) &0.0725&0.1860  \\ 
			(0, 0.5)&0.2101	&   0.77\\ 
			(0.1,0.05)&0.015&0.0529\\ 
			(0.1,0.1)&0.036&0.1039\\ 
			(0.2,0.1)&0.022&0.0738 \\ 
			(0.2,0.15)&0.044&0.14 \\
			(0.3,0.15)&0.0242&0.080
		\end{tabular}
		\caption{\label{tab:gap}Table showing the temperatures corresponding to the extrema $T^{*}$ of the isomagnetic plots of $\kappa_{xy}/T$ in Figs 2 and 3 of manuscript , together with the bulk gap $\Delta_{\text{bulk}}.$}
	\end{ruledtabular}
\end{table}

\section{Thermal Hall current decay in the bulk}
\begin{figure}
	\includegraphics[width=1\linewidth]{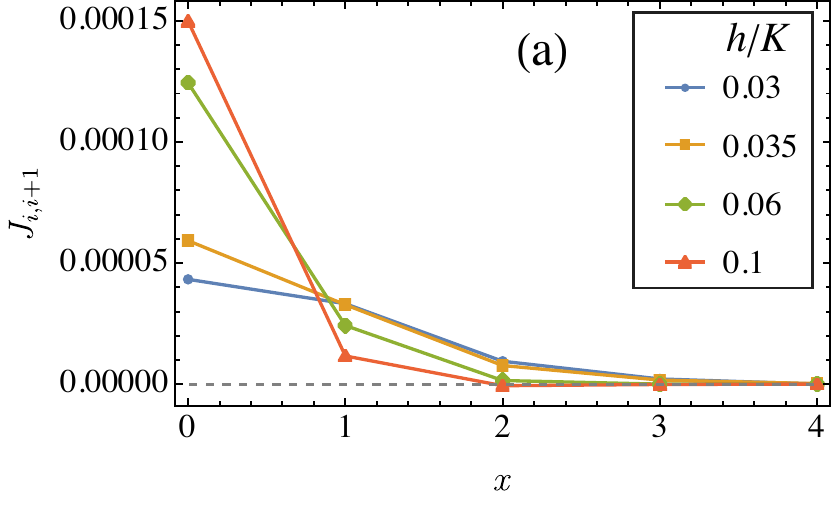}
	\includegraphics[width=1\linewidth]{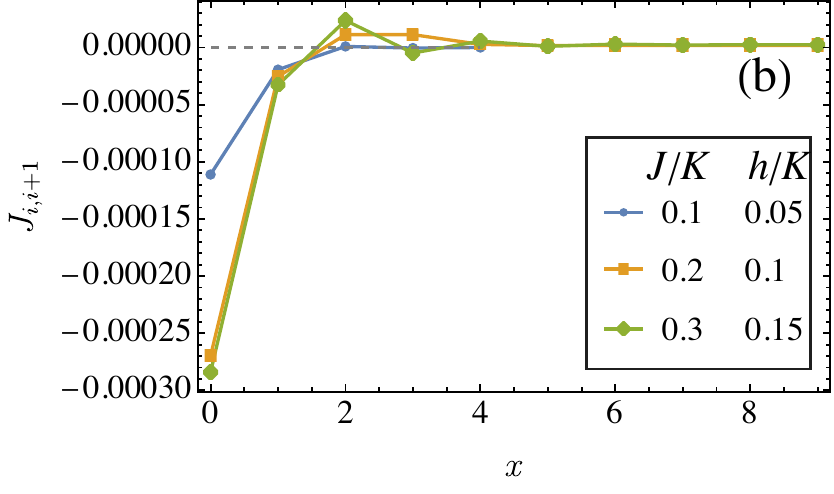}
	\caption{\label{fig:J_K_current}Plot (a) shows the decay of thermal Hall current along the the length of cylinder which is measured at the peak of  $\kappa_{xy}/T$ in pure Kitaev limit with varying magnetic field strength. We show the current decay up to half of the chosen cylinder length $(L_x=10)$ and in other half current will be equal and opposite. It is evident from the plot that thermal current vanishes deep in the bulk.  
		Plot (b) show the current decay in field suppressed SDW phase. Here, we choose the cylinder length $(L_x=20)$ to make sure current decays well with in bulk gap. }
\end{figure}

In gapped phases, the low-temperature behavior of thermal Hall effect is governed by low-lying edge modes. Figure \ref{fig:J_K_current} shows the decay of the thermal Hall current along the $x$-direction (cylinder axis) for in the Kitaev limit for various values of the Zeeman field along the $(111)$ direction. The current decays rapidly as we go deeper in the bulk allowing us to treat the left and right edge currents independently.
Figure \ref{fig:J_K_current} shows the current decay in the $J$-$K$ Kitaev-Heisenberg model in the presence of a Zeeman field along the $(111)$ direction. The ratios $J/K$ span both sides of the spin liquid ($J/K=0.1$) and stripy SDW phase ($J/K=0.2,\,0.3,$) boundary, and in the SDW phase, the field is chosen to just suppress SDW order and partially revive topological order (as measured by the topological entanglement entropy $\gamma$, see Fig. \ref{fig:gamma}). We  observe that the current changes its sign relative to the Kitaev limit and also decays rapidly into the bulk.\\
\\
\begin{figure}[h!]
	\includegraphics[width=1\linewidth]{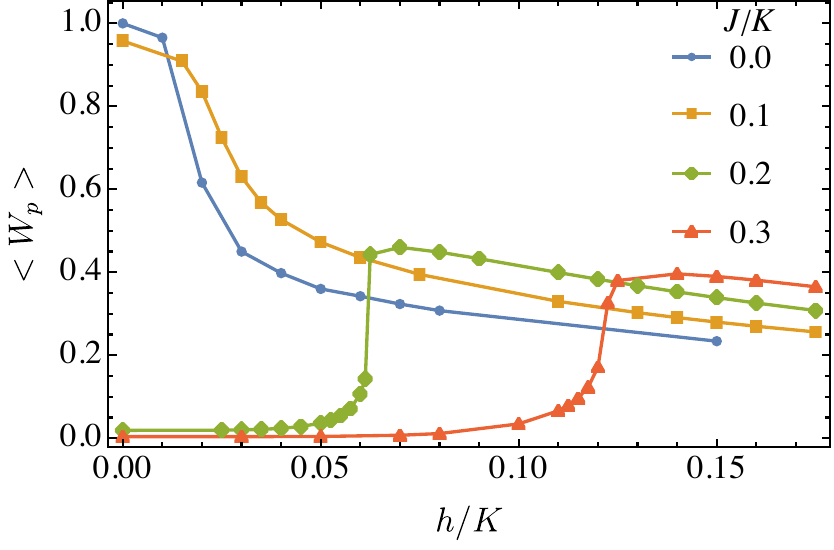}
	\caption{\label{fig:flux_revival}Ground state plaquette flux revival upon suppression of magnetic order. We considered the system size of $N_x=10, N_y=8,$ where ground state is computed using finite DMRG method.}
\end{figure}
\begin{figure}
	\includegraphics[width=1\linewidth]{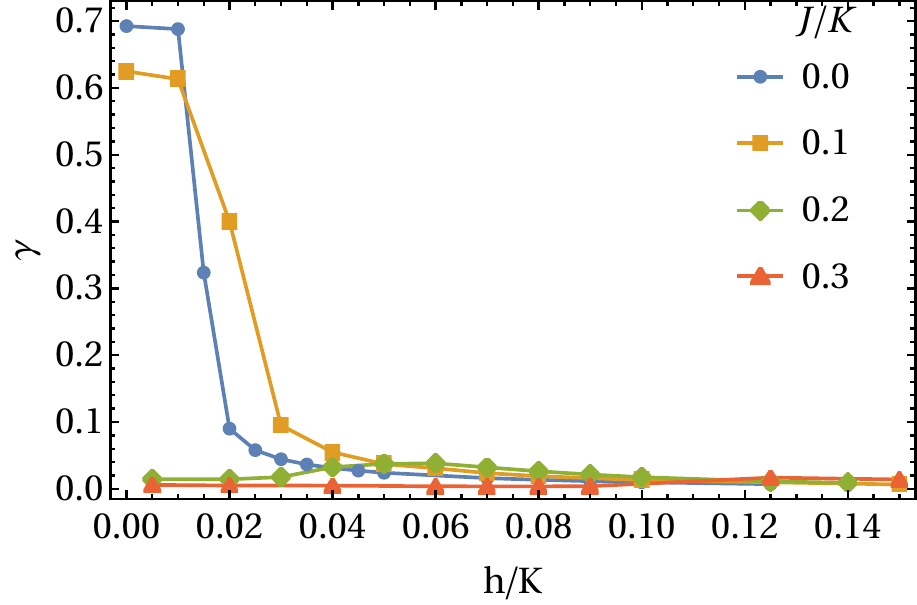}
	\caption{\label{fig:gamma}Field dependence of topological entanglement entropy $\gamma.$ It is small in the magnetically ordered phase as well as at high fields but there is a small, partial revival at intermediate fields where magnetic field is comparable to magnetic ordering scale. }
\end{figure}
\section{Flux revival and Topological entanglement entropy ($\gamma$) in field suppressed SDW phase}
Motivated by experimentally observed large thermal Hall effect near the onset of field suppressed SDW, we compute the plaquette flux expectation value in ground state for J-K model with varying field strength. We see that flux gets partially revived at field strength $\sim J/3$ - see figure  \ref{fig:flux_revival}.
We also compute topological entanglement entropy $\gamma$  using the Kitaev and Preskill construction \cite{kitaev2006topological}. In this intermediate field regime, we see that the  $\gamma(h)$ shows a weak revival, see Figure \ref{fig:gamma}, but stays way below $\gamma=\ln 2$ expected for ITO. We carried out this ground state computation with system size of $N_x=10, N_y=8$ using finite DMRG method.

\section{Spin - wave theory formalism}
We describe briefly the our spin-wave analysis for the thermal Hall response. The spin-waves in the field-polarized phase are obtained from Ref. \cite{Joshi_prb_magnon}. The Hamiltonian $H$ in the basis of spin wave Bloch states $\psi_{\kk}$ is
\begin{equation}
	\label{eq:h2k}
	H=\frac{S}{2} \sum_{\kk} \psi^{\dagger}_{\kk} M_{k} \psi_{\kk} \,,
\end{equation}

\begin{equation}
	\label{eq:mk_gen}
	M_{\kk} =
	\left(
	\begin{matrix}
		A_{\kk} & B_{\kk} \\
		B^{\dagger}_{\kk} & A^{T}_{-\kk}
	\end{matrix}
	\right) \,,
\end{equation}
where $k_{1,2} = a(\pm\frac{k_x}{2}+\frac{k_y \sqrt{3}}{2})$ with $k_x, k_y$ being the lattice momentum vectors. The bond length, $a,$ is henceforth set to unity. The matrices $A_{\kk}$ and $B_{\kk}$ are presented below in Eqs. \ref{eqn:Ak} and \ref{eqn:Bk}.

Following Refs. \cite{kapustin_thermal, guo_thermal}, the thermal Hall response (correctly accounting for the effect of bound energy magnetization currents) is given by
\begin{equation}
	\frac{d(\kappa_{xy}/T)}{dT}=-\frac{1}{2 T^3}\int \frac{d^dk}{(2 \pi)^3}\sum_{i}n_B^{\prime}(\mathcal{E}_{ik})\mathcal{E}^3_{ik}\Omega^z_{ik}
	\label{eqn:thermal_spin}
\end{equation}
Here $n'_B$ denotes energy derivative of the magnon (Bose) distribution function, $\mathcal{E}$ denotes the disperson of magnons and $\Omega^z_{ik}$ corresponds to Berry curvature at momentum $\kk.$
\begin{widetext}
	\begin{equation}
		A_{k}=\left(
		\begin{array}{cc}
			1+\frac{h}{S} & \frac{1}{3} \left(-1-e^{-i \text{k1}}-e^{-i \text{k2}}\right) \\
			\frac{1}{3} \left(-1-e^{i \text{k1}}-e^{i \text{k2}}\right) & 1+\frac{h}{S} \\
		\end{array}
		\right)
		\label{eqn:Ak}
	\end{equation}
	\begin{equation}
		B_k=\left(
		\begin{array}{cc}
			0 & \frac{1}{3} \left(-1-e^{-i \left(\text{k2}-\frac{2 \pi }{3}\right)}-e^{-i \left(\text{k1}+\frac{2 \pi }{3}\right)}\right) \\
			\frac{1}{3} \left(-1-e^{-i \left(-\text{k2}-\frac{2 \pi }{3}\right)}-e^{-i \left(-\text{k1}+\frac{2 \pi }{3}\right)}\right) & 0 \\
		\end{array}
		\right)
		\label{eqn:Bk}
	\end{equation}
\end{widetext}
%
\end{document}